# THERMODYNAMIC APPROACH TO LIQUID-TO-GLASS TRANSFORMATION AS AN ARREST TRANSITION IN POLYDISPERSE SOLUTION

Vladimir Belostotsky

Kurnakov Institute of General and Inorganic Chemistry, RAS, Moscow, Russia*

Thermodynamic multi-component solution solidification approach to liquid-to-glass transition is proposed and actual mechanisms underlying vitrification, other than viscous slowdown, are identified. Due to polydisperse aggregation in liquid state, glass-forming liquids, irrespective of chemical composition, appear to be mixtures of various quasi-components whose thermodynamic quantities shall be expressed not in terms of molar concentrations of actual chemical components, but in terms of relative concentrations of dominant structural units. Thermodynamically, any glass-former is expected to behave as multi-component solution and solidify in continuous temperature range between apparent liquidus and solidus temperatures that can be identified as glass-transition range. Using extended irreversible thermodynamics of polydisperse solutions it is demonstrated that upon quenching, diffusional and Brownian mass transport in such solutions is negated within heat removal timescale, which results in dynamical arrest of nucleation and growth in clusters and solid-liquid phase separation. Rapid solution solidification proceeds via successive cluster freezing in continuous temperature range, in line with cluster size dispersity, which can be described in terms of percolation in static polydisperse fractal ensemble where glass transition temperature naturally emerges as percolation threshold. Multi-component solution solidification framework is shown to be reconcilable qualitatively and quantitatively with the Mode Coupling - Random First Order Transition scenario. Finally, it is demonstrated that liquid-to-glass transformation is thermodynamic liquid-to-solid phase transition, glassy state of matter appears to be solid supersaturated solution of defects in otherwise perfect matrix, and the true equilibrium structure which glass is unable access is a crystalline one.

# I. INTRODUCTION

For almost a century since papers of Vogel [1], Fulcher [2], and Tamann and Hesse [3] were published, the phenomena pertaining to liquid-to-glass transition remain a very active area of exploration. A plethora of creative and seemingly rational but often incongruous and conflicting ideas as well as increasingly sophisticated mathematical models that account for some of the most perplexing features of glassy behavior have been put forward for resolving the glass transition conundrum [4]. Undoubtedly, many of them have captured some pieces of veracity [4,5,6,7,8,9,10,11,12] and continue being debated.

Among various concepts of glass transition, the mode coupling theory (MCT) [13,14,15,16,17] and the random first order transition (RFOT) theory [18,19,20,21,22,23] that had been introduced in 1980's, received substantial degree of general acceptance at that time, and continue dominate the glass science landscape ever since.

The MCT is a truly first principle, free of any dependence on phenomenological assumptions, statistical mechanics theory. The main theory input is static structure factor, $S(k)$, which contains information on the static density distribution of the material and is obtainable from the scattering experiments [24]. The MCT accounts for glass transition as a bifurcation from ergodic to non-ergodic behavior at critical temperature, $T_C$, and proposes "cage effect" as a microscopic mechanism for vitrification, however the predicted critical temperature, $T_C$, turned out to be sufficiently higher than expected experimental glass transition temperature, $T_g$ [25,26]. It was suggested, therefore, that since critical behavior predicted by MCT is correct, MCT could be considered as a generic, Landau-like, theory of glass transition, where $T_C$ may be treated as an adjustable parameter and rescaled to $T_g$ [27,28]. Recently presented multi-component generalized MCT (GMCT) suggests that predictive power of the theory may be improved by taking into consideration ignored so far subtle but important features of species-dependent structural correlations in $S(k)$ [29].

The RFOT theory, a spin-glass-inspired framework, is directly related to MCT at high temperatures since the temperature of dynamical arrest, $T_d$, of RFOT is $T_C$ of MCT (and the Goldstein's crossover temperature, $T^*$, that separates a free-flow regime from activated viscosity in supercooled liquids [6,27,30]). According to RFOT scenario, liquid-to-glass transition is a two-stage transformation: starting at $T_d$ and below, ergodic liquid fractures into a mosaic of metastable cooperatively rearranging regions, or entropic droplets, or glassy clusters [22] or "glassites" by analogy with "crystallites" in polycrystalline solids with which positive configurational entropy is associated [31]; or, in other words, liquid is trapped within walls made of the same frozen liquid where either activated fluctuations of grain-boundaries between glassy clusters sweep larger and larger regions of the bulk; or this is a nucleation of small droplets within the bulk that grow and percolate. At static temperature, $T_K$, where configurational entropy vanishes (Kauzmann temperature), the system undergoes random first-order thermodynamic phase transition into "ideal glass" state [27,30,32].

The authors of RFOT theory assert that it provides a theoretical basis for explaining mathematically much of glassy phenomena from initial equilibrium liquid state to the final state of ideal glass [31,33]. They also admit, however, that the fundamental physics of processes driving glass formation remain elusive [34]. Since the physical meaning of $T_C$ for finite-dimensional, non-mean-field glassy systems remains unclear, even the strongest selling point of RFOT, the crossover between a high-temperature MCT regime and lower temperature activated (mosaic) regime is very poorly understood even at a phenomenological level [27]. Importantly, from MCT-RFOT scenario $T_g$ has no way to emerge besides the artificial rescaling of $T_C$ to $T_g$ within Landau-like approach. So, a reasonable question arises as to why the attempts to reconcile apparent superior first-principle mathematical model of glass transition with physics run into so great difficulties. Could it be a case that there is an issue with the paradigm on which physics of glass transition rests?

Thus far, most of glass transition theories including MCT and RFOT remain guided by largely empirical and phenomenological considerations that liquid-to-glass transition occurs within liquid state through a viscosity-driven transformation into supercooled liquid and then into completely frozen liquid that looks and behaves mechanically exactly as solid but has disordered structure [4,13,14,15,16,22,31,32,35,36,37,38,39,40,41]. Since no significant structural changes appear to accompany glass transition and because the transformation occurs over a continuous temperature range, it has been postulated that the liquid and glass belong to the same thermodynamic phase [42]. Glass is a liquid, which is too viscous to flow, therefore the theory of glassy state shall be a byproduct of a theory of liquid viscosity [6], and must be focused exclusively on the detail analysis of microscopic relaxation dynamics in glass-forming liquid that govern viscous slowdown. Consequently, the theory of thermodynamic phase transition is irrelevant to understanding the liquid-to-glass transformation because the latter is a standalone phenomenon where thermodynamics has no direct role to play.

It is highly questionable, however, if not very unlikely that at normal conditions it is possible to escape the thermodynamic liquid-to-solid phase transition and cool a liquid down to temperatures where 'thermodynamics tells us it should not exist' [25]. Thermodynamics is the very foundation of all other sciences, and as such has supremacy over other disciplines without regard to the system's complex dynamic structure [43]. None of previous studies have explicitly acknowledged (except Ref. 25) and attempted to face this problem of the viscous slowdown model. Therefore, the motivation behind this work is to reconcile the phenomenology of glass transition with thermodynamics by identifying actual physical mechanisms underlying the liquid-to-glass transformation, other than viscous slowdown.

The paper is organized as follows. In Sec. II, we demonstrate that due to polydisperse aggregation into clusters or aggregates in liquid state, glass-forming liquids, even "simple" $SiO_2$, Se, or $H_2O$, appear to be mixtures of various quasi-components. Thermodynamically, they expected to behave as multi-component solutions and solidify in continuous temperature range between apparent liquidus and solidus temperatures that can be reconciled with glass-transition range. In Sec. III, based on extended irreversible thermodynamics analysis, we demonstrate that upon rapid cooling, diffusional and Brownian mass transport in such solutions is negated within the heat removal timescale, which results in dynamical arrest of nucleation and growth in clusters and solid-liquid phase separation. Rapid solution solidification proceeds via successive cluster freezing: within liquid phase above $T_g$ and within solid phase below $T_g$, which can be described in terms of percolation in a static polydisperse fractal ensemble where $T_g$ naturally emerges as a percolation threshold. The comparison of predictions of this multi-component solution solidification (MSS) approach with MCT-RFOT scenario lets us suggest that MCT does not fail to predict the precise location of the glass transition, it just captures the onset of transition of different kind, and that MSS can be reconciled with MCT-RFOT scenario both qualitatively and quantitatively, so that $T_C/T_d$ - $T_K$ interval obtains the physical meaning of the solution-solidification/glass-transition range. Finally, in Sec. IV, we discuss vitrification phenomenon in terms of thermodynamic phase transition theory and show that it is a thermodynamic phase transition within MSS framework.

## II. MULTI-COMPONENT SOLUTION APPROACH TO GLASS-FORMING LIQUIDS

Aggregation is a widespread, universally observed fate process related to self-organization phenomena [44,45]. Theoretical, experimental and computational advances of past several decades have proven the complex structural organization of liquids in the form of aggregates or clusters (see e.g. [46] and references therein). The kinetics of aggregation in the most general form was first formulated in the classical rate theory of Smoluchowski and his successors [47,48,49]. Ben-Naim [50,51] was perhaps the first who developed the formal theoretical foundation for the mixture-model of liquids, and demonstrated that even one-component liquid may rigorously be viewed as a multi-component system, and many properties of such a "simple" fluid as water (which is also a glass-former [52,53,54,55]) can be interpreted only within multi-component solution approach. Water aggregates or clusters have been experimentally determined thereafter (see e.g. [56,57] and references therein). In kinetic model of reversible aggregation in liquids by Kilian *et al.* [58] aggregates or clusters are considered as defect-saturated dynamic subsystems wherein collective modes of motions are excited, and it is suggested that liquids are condensed matter systems where *de facto* all the molecular units are linked in clusters. By aggregation into frustration-limited finite-sized clusters, liquids optimize themselves by generating nonhomogeneous in space hierarchically organized structure [59].

In the MCT-RFOT theory, although the terms "aggregation" and "cluster" are not broadly used (besides earlier works, e.g. [20,22]), the idea is very much behind the notions of "activated effects", "dynamical heterogeneities" [60] and "locally stable states" at temperatures above $T_C/T_d$, and "mosaic of metastable cooperatively rearranging regions or entropic droplets", "mosaic of aperiodic crystals" and "glassites" between $T_C/T_d$ and $T_K$ [27,31,61]. The MCT's prediction of the cage effect, i.e. the confinement of particles in local cages formed by their neighboring particles which, in turn, are trapped in their respective cages preventing them from moving around [24] is, in fact, the precise account for the microscopic mechanism of particles' trapping in clusters.

Apparently, polydisperse aggregation of atoms or molecules into finite-sized clusters in liquids is a diffusion-controlled process driven by particle-particle interactions [62]. At higher temperatures thermal movement prevents stable aggregation, aggregation is kinetically limited and reversible, and clusters (if any) are small-sized and short-lived. On approaching the thermodynamic phase transition temperature, $T_m$, from the above, aggregation dynamics become progressively more stable. The RFOT theory predicts that activated effects, otherwise metastable aggregation, first appear at the onset temperature, $T_O$, well above $T_d$ [22,27,63] and, as molecular dynamics (MD) simulations suggest, even well above $T_m$ [64]. Recent advances in electron correlation spectroscopy [65] have permitted the direct experimental visualization of the aggregation in liquid glass-formers as spatial heterogeneities [60,66,67,68,69].

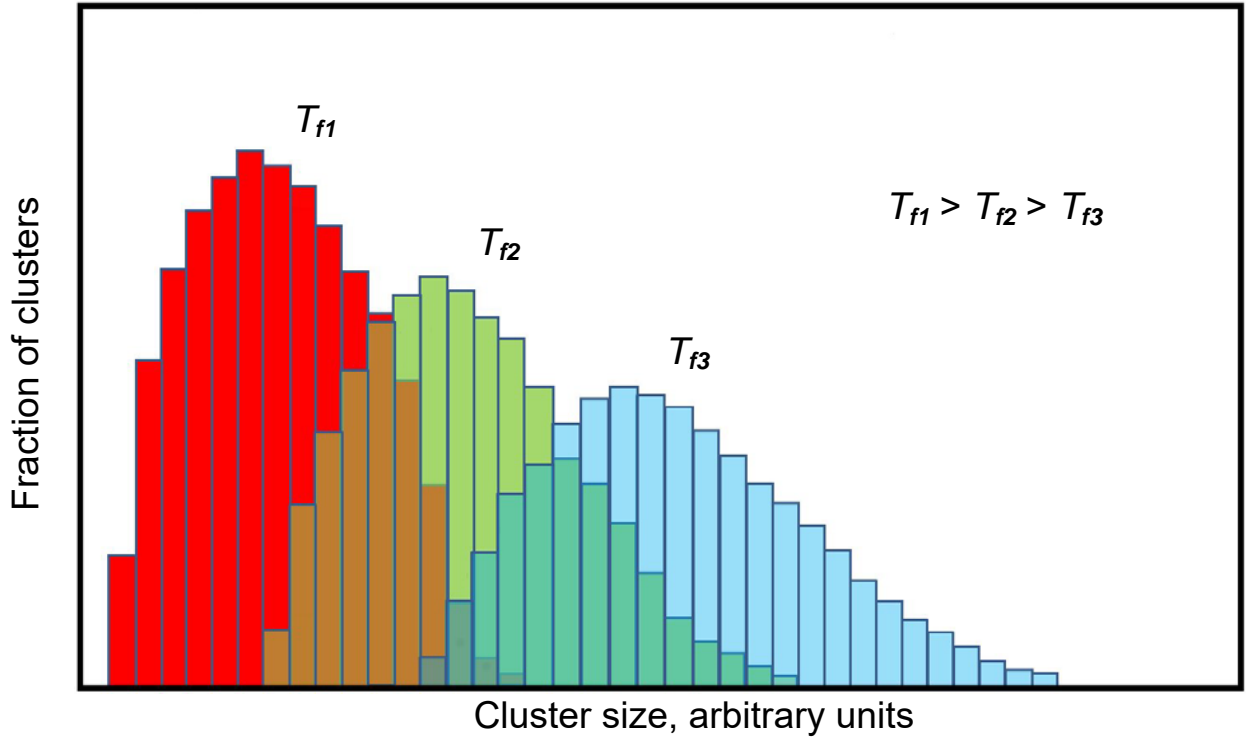

Fig. 1. Conjectured cluster size distribution in a glass former below the onset temperature of aggregation, $T_O$. The histograms represent the fractional concentration of clusters plotted against their sizes which characterize the composition of a glass-forming liquid as a polydisperse solution for three initial temperatures, $T_{f1}$, $T_{f2}$, and $T_{f3}$, from which a liquid is quenched to glassy state. Cluster size can be defined by a number of atomic or molecular particles it is comprised of, by volume, or by the effective length.

The cluster morphology, size dispersity and chemical reactivity depend on temperature of the system [70,71,72]. Temperature-dependent cluster size dispersity is controlled by Maxwell-Boltzmann's factor [58], as shown in Fig. 1. Clusters formed above or near $T_m$, exhibit liquid-like internal structure [58,73] and distinctly different freezing points depending on their size, density, shape and stoichiometry [74,75], therefore cluster morphology and size dispersity define the solution solidification range. We will return to this point in the next section.

On cooling across $T_m$, when the rate of heat removal is below critical, crystallization can be described as a two-stage process where aggregation into liquid-like clusters precedes the crystal nucleation which, in turn, proceeds via diffusional sorting of actual chemical components within clusters [76,77,78,79]. Clusters appear to be precursors to microcrystalline forms as liquid crystallizes on slow cooling: studies of morphologies of as-cast splat-cooled alloys show wide variety of grain sizes and shapes [80]. Evidently, crystal nucleation within clusters is one of the mechanisms that undergo dynamical arrest on rapid cooling which leads to freezing-in irregularities in the geometry of clusters' structure and shape [81]. On quenching, clusters may be considered relatively rigid and structurally stable to avoid shuffling of atomic particles into other configurations [25].

Aggregation into clusters in glass-forming liquids in the form of locally metastable species made up of at least several molecules has long been suspected [82,83,84,85,86,87,88,89]. Simmons and co-workers [90,91] were perhaps the first who determined that oxide glass-formers are composed of units having complex structure such as $(SiO_2)_6$, $(SiO_2)_8$, $(B_2O_3)_6$, etc rather than simple molecules like $SiO_2$ and $B_2O_3$, thermodynamically they behave as multi-component solutions, and their behavior can be described by the regular solution model. Likewise, Sactry *et al.* [74] described the aggregation in metallic glass-forming alloys as formation of complexes, so that binary systems behave as quasi-ternary regular solutions composed of unassociated atoms, say $A$ and $B$, and their complexes $A_pB_q$, where complexes may fail to nucleate if the corresponding metastable phase solidifies below $T_g$.

Following the findings of Simmons and co-workers [90,91] and Sactry *et al.* [74], the direct thermodynamic significance of polydisperse aggregation in liquids is that glass-formers, irrespective of actual chemical composition and, in particular, near $T_m$ and below, must be considered as multi-component solutions, whose thermodynamic quantities shall be expressed not in terms of molar concentrations of actual chemical components, but in terms of relative concentrations of dominant structural units [90]. As such, they are expected to solidify in continuous temperature range between apparent liquidus and solidus temperatures of the solution.

Any solution, irrespective of number of components, may be regarded as binary if its compositional variations are limited to a removal or addition of only one component [51,92]. Therefore, for further convenience and without loss of generality, we will treat any glass-forming liquid as a pseudo-binary solution. The fraction of species having structure closely resembling one of the precursor of embryonic nuclei we will view as a "perfect" solvent $A$, and the remainder comprised of assembly of species with various irregularities in shape, structure (including voids), and nonstoichiometry in composition will be considered as a "defective" solute $B$, and proportions of each are complimentary. The composition of such model binary solution can be written as $A_{1-C}B_C$ where $C$ is the concentration of solute. We can reasonably adopt that $A$ and $B$ are completely miscible in liquid phase and, as will be shown in the following, disregard any possible immiscibility issue in the transition interval and solid phase due to solute trapping in rapid solidification [93,94,95]. This liquid can be approximated by a regular solution model [91,96]. What distinguishes this liquid from typical regular solutions is that their components can be transformed one into another by addition or removal of defects. It is intuitively obvious that the fraction of defective components cannot be low, it ought to be at least 20-30%, otherwise crystallization becomes unavoidable [97,98,99,100].

## III. GLASS TRANSITION AS NON-EQUILIBRIUM SOLIDIFICATION OF MULTI-COMPONENT SOLUTION WITH DYNAMICAL ARREST OF NUCLEATION AND SOLID-LIQUID PHASE SEPARATION

In a large body of literature (mainly most recent) on rapid solidification processing of metal alloys, which is as non-equilibrium as liquid-to-glass transition, it has been demonstrated that thermodynamics and kinetics of liquid-to-solid phase transition are defined by the undercooling as the fundamental driving force of the transformation, and interplay between heat and mass transport fluxes that it causes and whose evolution involves different timescales (see [93,94,95,101,102,103,104,105,106,107,108,109,110,111] and references therein).

In a continuous system, under low heat removal speed, $V_T$, local thermal equilibrium is maintained both in the bulk liquid and at liquid-solid interface. The system evolution (i.e. solidification) is controlled by the bulk-liquid diffusive speed, $V_{Db}$, and interface diffusive speed, $V_{Di}$, while the interface velocity, $V_I$, is significantly lower than $V_{Db}$ and $V_{Di}$. Under these conditions (i.e. $V_T < V_I < V_D$), the component concentration field and temperature field are described by classical Fick and Fourier transport equations of parabolic type which implicitly assume infinite speed of heat and mass transport:

$$J = -D\nabla C$$

$$q = -\lambda\nabla T$$

where $J$ and $q$ are mass flux and heat flux, respectively, $D$ is diffusion coefficient, C is concentration, $\lambda$ is heat conductivity, and $T$ is absolute temperature. Phase transformation proceeds as first order thermodynamic transition via nucleation and growth, solid-liquid phase separation, and typically involves segregation of immiscible or partially miscible actual chemical components due to their different solubility in solid and liquid phases, all in accordance with equilibrium phase diagram.

Under large-amplitude perturbations, such as rapid liquid supercooling in amount of more than 0.5 $T_m$ before the onset of solidification, the classical irreversible thermodynamics (CIT) is no longer applicable due to violation of the local equilibrium hypothesis. Therefore, the local-nonequilibrium theory has to be used, specifically extended irreversible thermodynamics (EIT) [112,113], which includes dissipative fluxes, e.g. mass flux and heat flux, in the set of basic independent variables of the entropy.

For a binary system, the local nonequilibrium Gibbs equation can be written in the form [101,113]:

$$Tds = du + pd\upsilon - \mu dC - \frac{\tau_D}{D\rho^2}\left(\frac{\partial \mu}{\partial C}\right)JdJ - \frac{\tau_T}{\rho\lambda T}qdq \qquad (1)$$

in which $s$ is local specific entropy; $u$ is specific internal energy; $p$ is pressure; $\upsilon$ is specific volume; $\rho = \upsilon^{-1}$ is mass density; $C$ is concentration of component $B$; $\mu = \mu_A - \mu_B$, $\mu_A$ and $\mu_B$ are chemical potentials of components $A$ and $B$; $D$ is diffusion coefficient; $\lambda$ is heat conductivity; $\tau_D$ and $\tau_T$ are relaxation times to local equilibrium of mass (diffusion) flux $J$ and heat flux $q$, respectively, and $T$ is absolute temperature which retains local validity. Local nonequilibrium corrections to the classical Gibbs equation are represented in Eq. (1) by two last terms.

As shown in [106,113], the decay of $q$ in Eq. (1) for short-time heat transport beyond Fourier's law is governed by Maxwell–Cattaneo equation

$$q + \tau_T \frac{\partial q}{\partial t} = -\lambda \nabla T \qquad (2)$$

which, together with the energy conservation law

$$\rho c_\upsilon \frac{\partial T}{\partial t} = -\nabla q \qquad (3)$$

gives the hyperbolic heat conduction equation for the space-time evolution of the kinetic temperature $T$:

$$\frac{\partial T}{\partial t} + \tau_T \frac{\partial^2 T}{\partial t^2} = \frac{\lambda}{\rho c_\upsilon}\nabla^2 T \qquad (4)$$

where $c_\upsilon$ is the specific heat at constant volume. From Eq. (4), in turn, the velocity of heat propagation, $V_T$, can be obtained:

$$V_T = \left(\frac{\lambda}{\rho c_\upsilon \tau_T}\right)^{1/2} \qquad (5)$$

According to EIT [104,113], evolution of the mass (diffusion) flux in Eq. (1) is governed by the generalized Fick law:

$$\tau_D \frac{\partial J}{\partial t} = -(J + D\nabla C) \qquad (6)$$

which gives hyperbolic equations for the space-time evolution of mass flux and mass concentration:

$$\frac{\partial J}{\partial t} + \tau_D \frac{\partial^2 J}{\partial t^2} = D\nabla^2 J \qquad (7)$$

$$\frac{\partial C}{\partial t} + \tau_D \frac{\partial^2 C}{\partial t^2} = D\nabla^2 C \qquad (8)$$

Eqs. (6)-(8) predict the finite speed of mass transport, i.e. maximum velocity at which the diffusional perturbation can propagate in the liquid:

$$V_{Db} = (D/\tau_D)^{1/2} \qquad (9)$$

As shown in [104], sharp transition of solidification mechanism from coupled heat-mass transport to purely thermally-controlled regime occurs when the interface velocity $V_I$ and heat removal speed $V_T$ exceed the diffusion speed $V_{Db}$ of the most mobile actual chemical component. The decoupling of mechanisms can be described by the effective diffusion coefficient

$$D^* = D_b (1 - V_I^2 / V_{Db}^2) \qquad (10)$$

in which $D_b$ is the bulk-liquid diffusion coefficient of the most mobile actual chemical component. When $V_T \geq V_I > V_{Db}$, then $D^*=0$, which implies the switching to diffusionless solution solidification (see e.g. [93,94,95] and references therein). Glass transition in metallic alloys, for example, occurs at the heat removal speed, $V_T$, which, depending on actual material, can be of the order of several hundreds to thousand meters per second (see [94,104,108] and references therein) while characteristic diffusion speed, $V_{Db}$, does not exceed 10 m/s (typically ~ $10^{-1} - 10$ m/s [108]).

Diffusionless (chemically partitionless) solidification implies that the compositions of initial liquid and final solid phases are identical [93,110], so the system may be considered as behaving according to ideal solution. As already recalled above, nucleation and growth are diffusion-driven mechanisms, so in case of glass transition, diffusionless solidification implies the arrest of nucleation and growth of crystalline phase in clusters. Diffusionless solidification is the fundamental mechanism underlying the vitrification as a dynamical arrest transition.

As we turn now to a discrete, cluster-composed glass-forming liquid, we first have to specify the minimum space region to which local temperature can be assigned. We may reasonably assume here that temperature cannot vary within a cluster [106], therefore its own local temperature can be attributed to each cluster. As suggested in [103], since time to establish equilibrium between subsystems of discrete media could be greater than relaxation time to local equilibrium of each subsystem taken individually, upon quenching each cluster first relax to its local metastable equilibrium (i.e. freezing point and below) in a way which is nearly independent of the energy exchange between clusters, and then energy exchange between them occurs on the timescale $t \gg \tau_T$, supposedly in the form of fluctuation dissipation (i.e. according to fluctuation-dissipation theorem). As we pointed out above, depending on size, density, shape and stoichiometry, clusters exhibit distinctly different freezing points [74,75]. To put it simply, smaller, stoichiometric, clusters freeze first at higher temperatures, while the larger, defect-saturated clusters can maintain the heat for longer times and freeze at lower temperatures; so, to the first approximation, system evolves via successive clusters' freezing in continuous temperature range, in line with the size dispersity. As follows from Fig. 1, the majority of units in cluster ensemble should freeze within rather narrow temperature range corresponding to the most probable size in the cluster size distribution for each given temperature. The Fig. 1 explains the phenomenon of so-called "fictive temperature", i.e. the temperature of equilibrium liquid from which it is quenched to glassy state [114,115,116].

In a discrete, cluster-composed system, speed of nonvibrational thermal motion of structural units, which defines the timescale for phase separation, decreases roughly as reversed square root of average number of atomic particles in a unit [5]. Therefore, diffusionless solution solidification implies that within short heat flux dissipation timescale ($t \sim \tau_T << \tau_D$) the polydisperse cluster ensemble remains essentially motionless where solid-liquid phase separation mechanism has no time to intervene. This means that upon cooling, growing fraction of frozen, solid-like clusters remains dispersed and entrapped within static liquid domain spanning across the volume - until the system evolves from being predominantly liquid to predominantly solid, now with vanishing liquid fraction in the form of liquid-like clusters dispersed and entrapped within solid domain spanning across the volume - until freezing is fully complete. Or, in terms of MCT-RFOT theory, liquid is trapped within walls made of the same frozen liquid [27]. From this physical scenario, glass transition temperature, $T_g$, naturally emerges as the system's transformation threshold from being predominantly liquid to predominantly solid on cooling and the other way around on heating. Again, around $T_g$ the majority of units in cluster ensemble get frozen which corresponds to the most probable size in the cluster size distribution as shown in the Fig.1.

As we revisit the local nonequilibrium Gibbs equation (1), the arrest transition negates the terms related to mass transport (the third and the fourth variables of entropy) while the core equilibrium and heat flux variables of entropy remain intact:

$$Tds = du + pd\upsilon - \frac{\tau_T}{\rho\lambda T}qdq \tag{11}$$

This means that heat flux which is superimposed on the thermodynamic equilibrium, does not make the latter to negate, it only shifts experimentally observed macroscopic emergence of solid phase to the region of metastability. The particular physical significance of Eq. (11) then rests in the fact that thermodynamic phase transition is unquenchable and remains to underlie the vitrification. Therefore, for the model binary solution, kinetic phase diagram of the transformation can be constructed directly upon equilibrium phase diagram of ideal binary solution via extrapolation of kinetic phase boundaries as functions of temperature, composition, and cooling rate into regions of metastability [117,118], as shown on Fig. 2. Irrespective of cooling rate, equilibrium liquidus, that can be calculated as colligative freezing point depression of the solution, marks the onset of solidification (and glass transition) on microscopic level. The depression of freezing point of a solution depends on the fraction of solute, *C,* and is given by:

$$\Delta T(X) \cong (R T_m^2 / \Delta H_{mel}) \gamma_B C \qquad (12)$$

where $R$ is the universal gas constant, $\Delta H_{mel}$ is enthalpy of melting of the pure solvent, and $\gamma_B$ is the activity coefficient of solute [119]. As for the equilibrium solidus, it should mark the full completion of the solution solidification (and glass transition) on microscopic level.

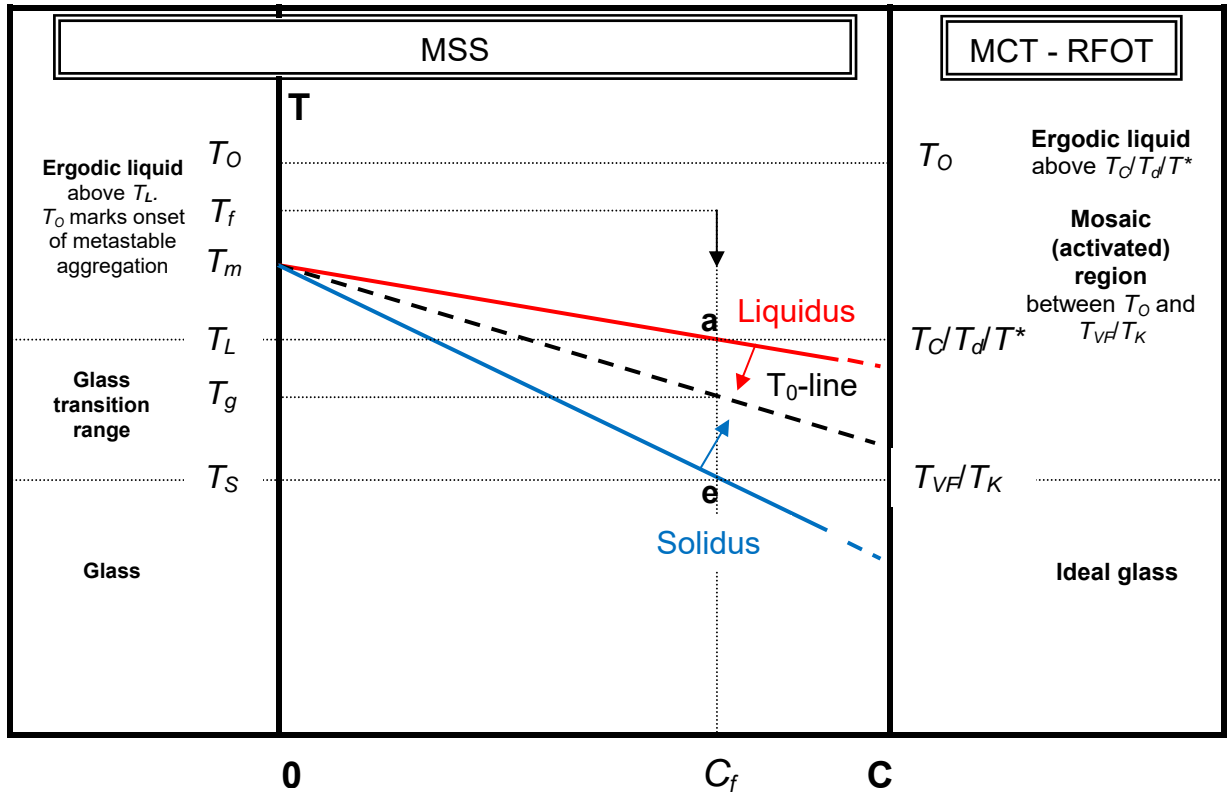

Fig. 2. Conjectured glass transition scenario within MSS framework with mapping to MCT-RFOT scenario. Schematic of a portion of the phase diagram shows liquid and solid phase equilibria expected for the model binary solution with composition, $C_f$, corresponding to initial temperature from which a liquid is quenched to glassy state, $T_f$, and kinetic single $T_0$-line to which kinetic liquidus and solidus curves converge upon quenching.

The outlined above MSS scenario of liquid-to-glass transformation as a dynamical arrest transition well corroborates with the metallurgists' experimental observations that, upon undercooling rate increase, the kinetic liquidus of alloy tends to drop while the kinetic solidus tends to raise [120,121]. At the supercritical heat removal speed of diffusionless solidification regime, kinetic liquidus and solidus curves converge into single "$T_0$-line" that corresponds to the temperature, $T_0$, at which the solid and liquid have the same composition and Gibbs free energy [120]. Under this cooling regime, a binary alloy behaves like a pure metal with freezing temperature $T_0$, similar to glass transition. Based on the foregoing, we may reasonably suggest here that the evolution of the phase diagram from equilibrium to the kinetic arrest-transition single $T_0$-line, actually visualize the growing time lag between the initial solid phase formation on microscopic level in the form of solid-like clusters dispersed within liquid domain and its macroscopic emergence as soon as the system transforms from being predominantly liquid to predominantly solid. The immediate conclusion that can be drawn from this observation is that $T_0$ and $T_g$ are intimately related if not the same. Metallurgists consider that vitrification of metallic alloys is achievable when undercooling exceeds $T_0$ [122].

Importantly, the shape of the phase diagram given in Fig. 2 and the position of isopleth which the solution follows on cooling are controlled by initial temperature, $T_f$, from which ergodic liquid is quenched to glassy state. The $T_f$ defines apparent composition of the glass-former as a polydisperse solution as shown in Fig.1, and thus sets its $T_L$ and $T_S$ which, in turn, specify the width of glass transition range. The only fixed parameter of the phase diagram is $T_m$. From Fig. 2 it also follows that the processes observed between $T_g$ and $T_S$ and described in the literature as "secondary relaxation" [123,124], take place, in fact, within glass transition range.

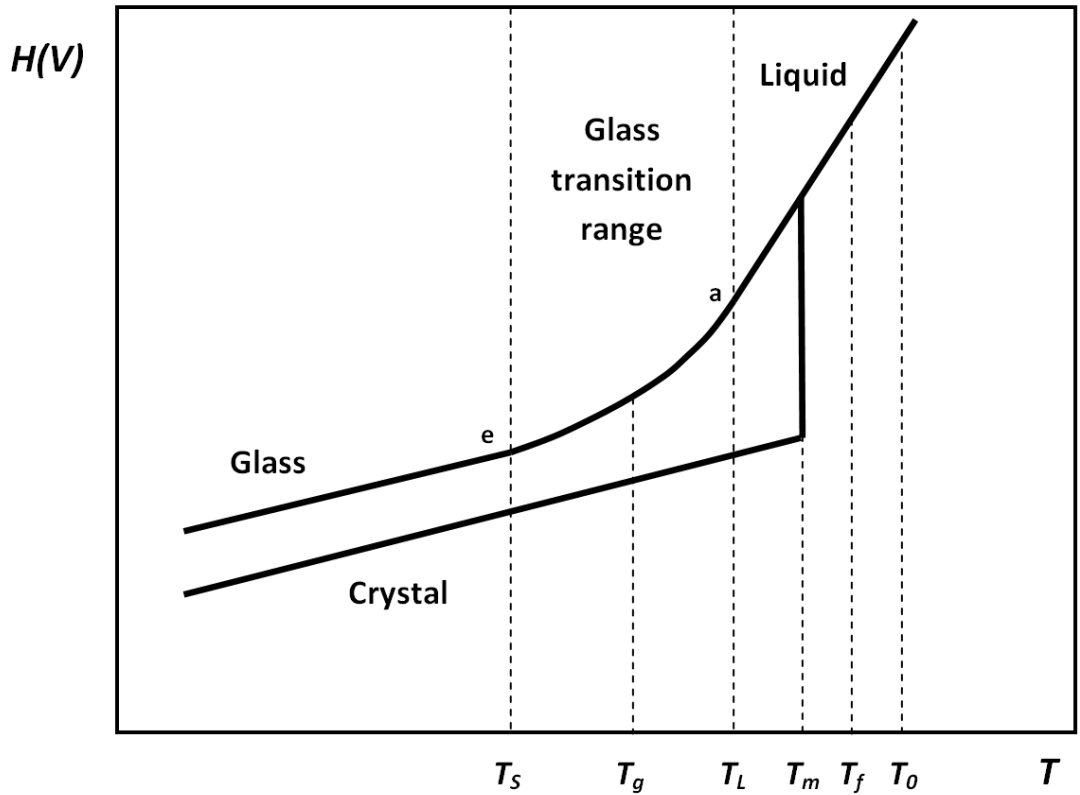

Fig. 3. Enthalpy (specific volume) as a function of temperature diagram for glass transition between $T_L$ and $T_S$ versus crystallization at $T_m$.

The nature of liquid-to-glass transition becomes even more apparent from enthalpy (or volume) vs. temperature diagram for glass transition given in Fig. 3. The comparison of Fig. 2 and Fig. 3 makes it evident that the glass transition range in Fig.3 corresponds to the solidification range between points **a** ($T_L$) and **e** ($T_S$) in Fig. 2. Between $T_m$ and $T_L$, the decrease of the melt's enthalpy (or specific volume) continues to follow the liquid line because the substance, which is described as "supercooled liquid", within this temperature range persists being, in fact, ergodic, freely flowing liquid. Below $T_S$, the substance which we call "glass" is completely solidified solution, and its further specific volume decrease with temperature is approximately parallel to that of the solid (crystal) line. This reflects the fact that glass is indeed solid, while its excess enthalpy (and specific volume) over crystal is directly related to enthalpy of mixing of perfect and defective components.

Between $T_L$ and $T_S$, the transition from liquid to glass is observed as a smooth curve connecting liquid and solid lines because it occurs as a continuous and gradual solidification with dynamically arrested liquid and solid phase separation. This is true glass transition range within the MSS framework.

A comparison of the outlined above MSS approach with the predictions of MCT-RFOT theory reveals the striking similarities between scenarios. As can be seen from Table 1, $T_C$ calculated for various glass-formers is found well above $T_g$, in the temperature range $(0.8\text{-}0.9)T_m$, which corresponds to the range of the depression of the solutions' freezing point, and marks the onset of glass transition as solution solidification on microscopic level within MSS framework. It is reasonable to suggest therefore that the MCT crossover temperature, $T_C$, can be mapped to $T_L$ which is also a crossover temperature where the system begins gradually falling out of equilibrium.

At the lower end of the transition range, static temperature of RFOT theory, $T_K$, seems to mark the completion of the transition. One can argue therefore that it may be mapped to $T_S$ of MSS scenario. However, $T_K$ is not a measurable or calculable temperature. In order to determine $T_K$, one has to extrapolate positive entropy of glass over crystal to zero [138] which, from our standpoint, seems to be an unwarranted fit since the former is directly related to the positive configurational entropy of mixing of perfect and defective components in glass that cannot be physically eliminated without crystallization. On the other hand, empirically $T_K$ is close to Vogel-Fulcher temperature, $T_{VF}$, at which the relaxation time appears to diverge reflecting elimination of non-vibrational modes of the liquid [125,139]. Thus, qualitatively and quantitatively $T_C/T_d - T_K/T_{VF}$ interval may obtain the physical meaning of the solution-solidification/glass-transition range. Table 1 entries specify glass transition ranges within MSS framework for four glass-formers: orthoterphenyl (184 K – 290 K, $T_g \sim 243$ K), salol (135 K – 270 K, $T_g \sim 218$ K), propylene carbonate (130 K – 192 K, $T_g \sim 160$ K), and toluene (103 K – 153 K, $T_g \sim 117$ K).

Table 1. Important temperatures: Kauzmann, $T_K$, Vogel-Fulcher, $T_{VF}$, glass transition, $T_g$, melting, $T_m$, MCT's critical, $T_C$, and their relations for various glass-formers

| Substance | $T_K$ [125] | $T_{VF}$ [125] | $T_g$, K | $T_m$, K | $T_g/T_m$ | $T_c$, K | $T_c/T_m$ | Method | Reference |
|---|---|---|---|---|---|---|---|---|---|
| $Pd_{43}Cu_{27}Ni_{10}P_{20}$ | | | 582 | 802 | 0.73 | 710 | 0.89 | Radiotracer diffusion | [126] |
| $Ca_{0.4}K_{0.6}(NO_3)_{1.4}$ | | | 333 | 483 | 0.69 | 376-380 | 0.86-0.87 | Light scattering | [127] |
| orthoterphenyl (OTP) | 200 | 184 | 243 | 329 | 0.74 | 290 | 0.88 | Light scattering | [128] |
| Phenyl salicylate (salol) | 167 | 135 | 218 | 315 | 0.69 | 256-270 251-261 | 0.81-0.86 0.80-0.83 | Neutron scattering Light scattering | [129] [133] |
| $Na_{0.5}Li_{0.5}PO_3$ | | | 515 | 749 | 0.68 | 601-637 | 0.80-0.85 | Neutron scattering | [130] |
| $n$-butylbenzene | | | 127 128 | 185 | 0.69 | 150 160 | 0.81 0.86 | Light scattering Solvation | [131] [134] |
| Propylene carbonate (PC) | 126 | 130 | 160 | 218 | 0.73 | 170-210 176 182-192 | 0.78-0.96 0.81 0.83-0.88 | Neutron scattering Solvation Light scattering | [132] [134] [136] |
| Toluene | 96 | 103-108 | 118 117 | 178 | 0.66 | 140-146 153 | 0.79-0.82 0.86 | Light scattering & Neutron scattering Light scattering | [135] [137] |

Around $T_g$ rounded discontinuities of the constant pressure heat capacity, $C_P$, the expansion coefficient, $\alpha_P$, and the isothermal compressibility, $K_T$, are observed which point out to the striking similarity between glass transition and thermodynamic phase transition [140,141]. It will be discussed separately in the next section. The discontinuities in properties or jumps of thermodynamic functions that related to phase transitions always occur across the phase interface. This means that growing solid phase, which is dynamically jammed and entrapped within the walls of liquid domain above $T_g$, near $T_g$ and below emerges macroscopically and begins dominating the character of the solidifying substance. In other words, at $T_g$ and below solid phase outgrows liquid one, overcomes the percolation threshold, and forms an infinite cluster. The remaining isolated liquid-like clusters are entrapped within the walls of solid phase and solidify completely as $T_S$ is reached. From this observation it can be suggested that between $T_L$ and $T_S$ fractal formalism is applicable to describe the dynamics of rapidly solidifying multi-component ensemble of clusters (or a mosaic of glass nodules or "glassites" in RFOT terminology [27]) which can be treated as "fractal elements". Detail consideration of this particular proposal, however, is beyond the scope of this work. It noteworthy, though, that interesting approach to this issue is found in works of untimely deceased Richard Wool [8,142,143].

From standpoint of thermodynamics, resulting frozen system, otherwise glass, appears to be a supersaturated multi-component solidified solution.

## IV. GLASS TRANSITION AS PHASE TRANSITION

The foregoing consideration has revealed the ambivalent nature of the glass-forming liquid approaching $T_m$ from the above, and cooling rate appears to be the critical parameter defining how the system will respond thermodynamically to the perturbation, i.e. the transformation route the system will follow on cooling across $T_m$.

On slow cooling, the diffusional sorting of actual chemical components in liquid ahead of advancing solid-liquid interface leads to elimination of defective components by way of ordering and crystallization of the perfect one. Liquid-solid transformation occurs as first-order phase transition through nucleation and growth with formation of the macroscopic phase interface, which in turn is the cause of jumps in all extensive thermodynamic quantities accompanying the transition from one phase to another. For the first-order phase transition $\Delta V_{PTI} \neq 0$, $\Delta S_{PTI} \neq 0$, thus phase transition heat $Q_{PTI} = T \, \Delta S_{PTI} \neq 0$, and the derivative $(dP/dT)_{PTI} \neq 0$ for each condensed phase.

Glass transition, the route the system follows on quenching, is the transformation for which volume and entropy change smoothly. However, glass transition exhibits all formal qualitative features of the second-order (continuous) phase transition [144,145]: observed changes in physical properties are enormous (e.g., static shear modulus value changes from a liquid's zero to that of a crystal); they occur in wide temperature interval without formation or disappearance of phase interfaces. At the same time, jumps are observed in the constant pressure heat capacity, $C_P$, the expansion coefficient, $\alpha_P$, and the isothermal compressibility, $K_T$, which are first derivatives of those quantities that have jumps in first-order phase transition:

$$C_P \equiv \left(\frac{\partial H}{\partial T}\right)_P ; \qquad \alpha_P \equiv \frac{1}{V}\left(\frac{\partial V}{\partial T}\right)_P ; \qquad K_T \equiv -\frac{1}{V}\left(\frac{\partial V}{\partial P}\right)_T \qquad (13)$$

In literature, the main reasoning against considering phase transition in a sense of Ehrenfest as the fundamental critical phenomenon underlying glass transition is centered around the non-equilibrium nature of glass transition and the fact that glass transition temperature and the width of transformation range depend on fictive temperature and cooling rate [140]. Besides, it is argued that the formal examination of jumps in $C_P$, $\alpha_P$, and $K_T$ with Prigogine–Defay ratio (PDR) [25,146,147] seems discouraging as well for classifying glass transition as second-order phase transition.

The PDR has been deduced from the Ehrenfest equations [148]:

$$\left(\frac{dP}{dT}\right)_{PT2} = \frac{1}{T_{PT2}V}\frac{\Delta C_P(T_{PT2})}{\Delta \alpha_P(T_{PT2})} \qquad (14)$$

$$\left(\frac{dP}{dT}\right)_{PT2} = \frac{\Delta \alpha_P(T_{PT2})}{\Delta K_T(T_{PT2})} \qquad (15)$$

where $T_{PT2}$ is the temperature of the second-order phase transition. Both equations combined yield desired PDR ($\Pi$) that equals unity at second-order phase transition:

$$\Pi \equiv \frac{\Delta C_P(T_{PT2})\Delta K_T(T_{PT2})}{T_{PT2}V[\Delta \alpha_P(T_{PT2})]^2} = 1 \qquad (16)$$

The formal application of PDR to glass transition with $T_g$ substituted for $T_{PT2}$ converts Eq. (16) into inequality because of violation of the second Ehrenfest equation (15) [140]. Typically, PDR calculated for glass transition is greater than unity and varies in the range between 2 and 5 [140,146]. For extreme case of vitreous silica, PDR is greater than 10,000 [152]. It is argued that the fact that PDR>1 cannot be explained solely by uncertainty in measurements of $T_g$ and lack of sharpness of discontinues of thermodynamic quantities; it is believed to be an evidence that complete description of glass transition requires more than one order parameters [149] or internal parameters [150] including fictive temperature, $T_f$, and fictive pressure, $P_f$. The detail review of the application of PDR to thermodynamic analysis of glass transition is found in Ref. 151.

From standpoint of MSS framework, however, volatility of $T_g$ and the width of transformation interval, their dependence on $T_f$ and cooling rate, owe this behavior to the defining effect of $T_f$ and cooling rate on actual composition of the system as multi-component solution. In addition, it is worth re-emphasizing that $T_g$ is not the actual transformation temperature because at $T_g$ the system is still in the middle of transition. Therefore, usage of $T_g$ as a substitute for phase transition temperature, $T_{PT2}$, in the PDR [25,147,152] appears to be an unwarranted fit. Moreover, it is doubtful whether such a formal examination (requirement that PDR=1) is applicable to multi-component solutions whose solidification occurs over the temperature interval: Gupta and Haus [150] have shown that even with single internal parameter a system would always have PDR>1 provided the system is multi-component.

As for the non-equilibrium nature of glass transition, in addition to what has already been discussed on this matter in previous section, it is instructive to recollect over again that rapid solidification processing of crystalline metal alloys is as non-equilibrium as liquid-to-glass transformation. The cooling rate increase (provided it is below critical threshold) leads to greater microcrystalline structure refinement [80,153] without altering thermodynamics of the processing as first-order phase transition. This pattern of behavior persists until the critical heat removal speed is reached which marks the switch to diffusionless solidification as defined by Eq. (10) beyond which grain boundaries that serve as unsaturable sink become unreachable for defects due to their entrapment. This does not mean that the system avoids phase transition. Sobolev has argued in [105] that the transition from diffusion-limited to diffusionless solidification is itself a critical phenomenon, and we must agree with this assessment. From standpoint of MSS framework, this is indeed a first-to-second-order phase-transformation crossover. Therefore, amorphous metallic alloys are as solid as crystalline even though crystalline forms are absent or undetectable with available methods. Again, glass is not a liquid, which is too viscous to flow, it is a solid supersaturated solution of defects [153] in otherwise perfect matrix, and the true equilibrium structure which glass is unable access is a crystalline one. (Actually, from the standpoint of thermodynamics, all solids containing defects are solid solutions, but it is typically ignored). This definition of glass becomes even more evident when we recollect that crystalline solids may be vitrified via direct injection of defects by, e.g., irradiation [154]: after supercritical irradiation dose and subsequent annealing, their structure becomes remarkably close to that of corresponding glass.

**CONCLUDING REMARKS**

This work offers a unified, consistent, and coherent approach to the problem of liquid-to-glass transition within the framework of extended irreversible thermodynamics of polydisperse solutions which is valid for all types of glass formers. It demonstrates that glass transition is not merely kinetic or thermodynamic phenomenon but rather interplay between thermodynamics and kinetics where kinetics defines thermodynamics of the system's transformation route from liquid to solid state. It upholds the qualitative and quantitative validity of the equations of the MCT-RFOT theory which accounts for glass transition mathematically, so it can be interpreted physically based upon MSS framework.

**ACKNOWLEDGEMENT**

The author is deeply grateful to Dr. Sci., Prof. Sergei Aristarkhovich Dembovsky for inspiration of this work, and dedicates this paper to his memory.

_______________________________

* Retired, email: vladbel@erols.com